# The Digital Divide in Process Safety: Quantitative Risk Analysis of Human-AI Collaboration

He Wen
Department of Civil and Environmental Engineering,
University of Alberta, Edmonton, AB T6G 2E1, Canada

**Abstract**

Digital technologies have dramatically accelerated the digital transformation in process industries, boosted new industrial applications, upgraded the production system, and enhanced operational efficiency. In contrast, the challenges and gaps between human and artificial intelligence (AI) have become more and more prominent, whereas the digital divide in process safety is aggregating. The study attempts to address the following questions: (i)What is AI in the process safety context? (ii)What is the difference between AI and humans in process safety? (iii)How do AI and humans collaborate in process safety? (iv)What are the challenges and gaps in human-AI collaboration? (v)How to quantify the risk of human-AI collaboration in process safety? Qualitative risk analysis based on brainstorming and literature review, and quantitative risk analysis based on layer of protection analysis (LOPA) and Bayesian network (BN), were applied to explore and model. The importance of human reliability should be stressed in the digital age, not usually to increase the reliability of AI, and human-centered AI design in process safety needs to be propagated.

**Keywords**：process safety, human, AI, quantitative risk analysis, digitalization

## 1. Introduction

Information and communication technologies (ICT), or digital technologies, such as artificial intelligence (AI), 5G, cloud computing, big data, Internet of Things, mobile internet, show a geometric growth in technological development and industrial application. The digital economy was estimated to range from 4.5% to 15.5% of world GDP (UNCTAD, 2019). Digitalization has tremendously changed the world, society, economy, industry, manufacturing, process industry, and even in process safety (Lee et al., 2019), occupational safety (European Agency for Safety & Health at Work [EU-OSHA], 2018), bringing numerous business chances, potential increase, and new digital threats (Chen et al., 2020; Ghobakhloo, 2020; Ivanov et al., 2019).

One of the challenges is the rising risk in human-AI interaction (Briken, 2020). However, what is AI in the process safety context? AI is the typically integrated technology in digitalization. Researchers have different opinions on how to accurately define AI, while a consensus is that "intelligence" is the crucial difference between human beings and animals, machines, in terms of cognitive ability, and AI is the attempt to reproduce this ability in computer systems (Wang, 2008). The definition of AI in some ISO/IEC standards (ISO/IEC TR 24029-1:2021, ISO/IEC TR 24028:2020, ISO / IEC TR 29119-11: 2020), is the capability of an engineered system to acquire, process and apply knowledge and skills. Accordingly, the control system in process industries, such as the basic process control system (BPCS), alarm system (AS), safety instrumented system (SIS), and emergency shutdown (ESD), should be the applications of AI. Figure 1 shows the AI in layer of protection analysis (LOPA).

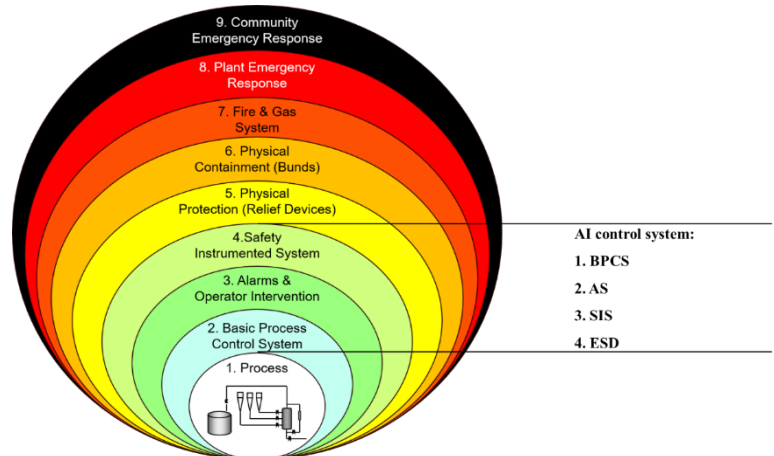

Figure 1: LOPA and AI.

While it is to be seen that the significant process accidents have not been reduced (Khan et al., 2021), it seems that the continuous improvement of automation and digitization cannot eliminate the risk. Yet, the problem would be the reliability of AI control system (Benson et al., 2021), or the human reliability (Khan et al., 2021).

The AI control system failure research includes the component reliability of the senor, logic solver, and actuator, and system reliability in specific applications. Since the 1980s, sensor reliability has entered the researcher's field of vision, and statistical methods by manufacturers and industry users were the critical data sources. One of the leading is the Offshore Reliability Data Handbook (OREDA) since 1981, to collect reliability data for safety equipment (OREDA, 2021), and similar handbooks, such as Reliability Data for Safety Equipment (PDS Data Handbook) (SINTEF, 2021), and some reliability handbooks by Center for Chemical Process Safety (CCPS). In 1996, Idaho National Engineering Laboratory studied the failure modes and average failure rates of temperature, pressure, flow, and level sensors (Cadwallader, 1996). The analysis was from sensor structure and function level, with failure mode and effect analysis (FMEA), and benchmarked different sensor reliability data sources.

With the technology upgrade and advanced packaging of sensors, especially the Micro-Electro-Mechanical System (MEMS) sensors, the precision and reliability of sensors have been greatly improved (McCluskey, 2002; Tadigadapa & Najafi, 2001). For this reason, the research on AI reliability switched from the component level to the system application level. As SIS required higher reliability than BPCS, the reliability of SIS with safety integrity level (SIL) calculation became the focus in different process industries gradually (Alizadeh & Sriramula, 2018; Chang et al., 2015; Liu & Rausand, 2013; Nasimi & Gabbar, 2016), and one common typical application is LOPA (CCPS, 1993; Dowell, 1998). Currently, the functional safety in process industries has been defined in IEC 61508, ISO TR 12489, and industrial handbooks, for example, the hardware fault tolerance (HFT) design. The problem is that the current research focuses on common cause failure and mechanical failure, while there is not enough attention on random disturbance, system integration failure, and model uncertainty.

Besides AI reliability, human reliability is another aspect influencing the digitalization impact on process safety. Human error or human factor research started in France in the 1930s (Hollnagel, 2018), when the journal *Le travail humain* published the first volume in 1933 by Presses Universitaires de France. In the UK, the Ergonomics Research Society was founded in 1946. In the US and some other countries, human error had also been considered in the mid-1940s. However, human error had not drawn much attention until the Three Mile Island (TMI) nuclear power plant accident in 1979. Even before that, in 1975, the WASH-1400 Reactor Safety Study by the US Nuclear Regulatory Commission (USNRC, 1975), which is the milestone of quantitative risk analysis (QRA), had involved the error of operator action in the event tree analysis (ETA), and the analysis group attempted to use human error rates. Since then, quantitative methods were developed for human error probability (HEP) or human reliability assessment (HRA) (Park et al., 2019; Riccardo et al., 2020). Task-based, cognition-based, and reliability-based methods were developed (Kirwan, 1998). US Nuclear Regulatory Commission played a vital role in this research area, as they developed several practical techniques (Huang et al., 2017)(Forester et al., 2006), such as technique for human error rate prediction (THERP) (1983), success likelihood index method (SLIM) (1984), accident sequence evaluation program HRA procedure (ASEP) (1987), a technique for human error analysis (ATHEANA) (1999), standardized plant analysis risk human reliability analysis (SPAR-H) (2005), and an integrated human event analysis system for nuclear power plant (IDHEAS) (2017). In the meantime, they accumulated one of three famous HEP databases, Scenario Authoring, Characterization and Debriefing Application (SACADA), by nuclear plant simulators. Currently, the methods tend to mix techniques, primarily based on cognition and task, and it is the original thought of cognitive reliability and error analysis method (CREAM) (Hollnagel, 1998). The problem is that HEP research is usually separated from AI failure, as researchers focus on the specific scenarios for human error and factor study. Though some QRA methods, such as LOPA, consider human intervention to deal with the alarm simply without extension on HEP analysis (Baybutt, 2002; Dowell, 1998; Myers, 2013).

Nevertheless, as AI spreads in industry applications, the interaction between humans and AI becomes common, and collaboration safety problems increase dramatically. At the same time, the combination of AI failure and human error has not drawn enough attention. One combination is the human in the loop control system, which is to design the human as a component in the closed-loop control system (Cohen & Singer, 2021; Edwards & Lees, 1971), or to study the human intervention and its influence (Wu et al., 2021). To quantify the risk between humans and AI, the first to do is to study the collaboration relation. In some industrial applications, autonomous vehicles (Aptiv et al., 2019), collaborative robots (Aaltonen et al., 2018; Magrini et al., 2020), and medical devices (Freschi et al., 2013) showed leading-edge exploration. The problem is that the types of human-AI collaboration in process safety are still not identified, and the risk of human-AI collaboration needs to be quantified.

Specifically, what the current research lacks are: 1) the conjoint analysis of AI failure and human error; 2) the quantitative analysis of human error in other process industries besides of nuclear industry; 3) how to deal with the data uncertainty in the probabilities of AI failure and human error; 4) how to quantify the risk in human-AI collaboration. Significantly, the gap between humans and AI is a kind of digital divide, raising more potential risks in the digital era.

For the above reasons, this research aims to set up a model to illustrate the collaboration between humans and AI, quantify the risk of human-AI interaction, and discover the vital components and actions.

## 2. Methodology

### 2.1. Research framework

This research aimed to achieve the quantitative risk analysis of human-AI collaboration (Figure 2), which was the joint area of human error and AI failure based on data analysis. The research started with the literature review and brainstorming to address the prepositive questions: 1) What are the types of AI in process safety? 2) What is the difference between AI and human in process safety? 3) How do AI and human collaborate in process safety?

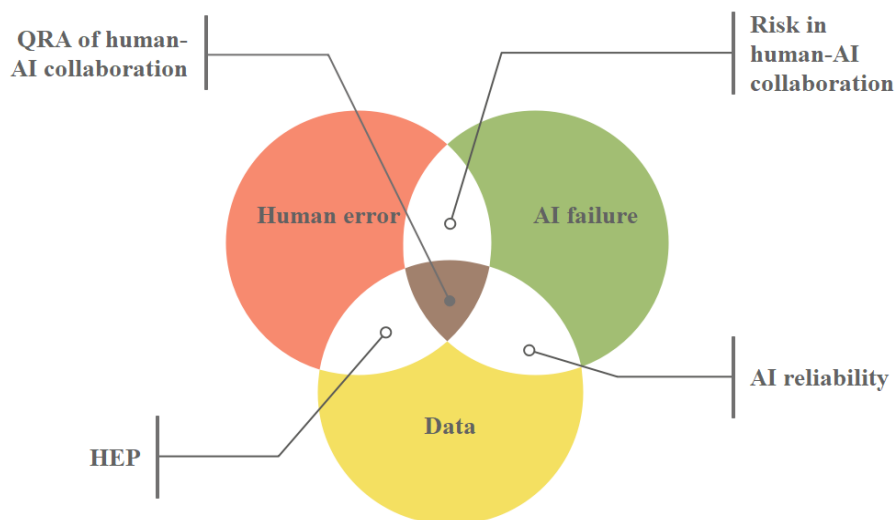

Figure 2: The relation of human error, AI failure, and data.

After that, a complete qualitive risk analysis was conducted to identify and summarize the challenges and gaps in human-AI collaboration. Moreover, QRA was applied with LOPA and BN to specify and assess the risk in human-AI collaboration. In this step, the probabilities of human error and AI failure were accessed from LOPA practices, IDHEAS, and industry application practices. In addition, a published case was introduced (Thepmanee, 2018), and a comparative case study with different scenarios was implemented. The research followed the procedure defined in Figure 3.

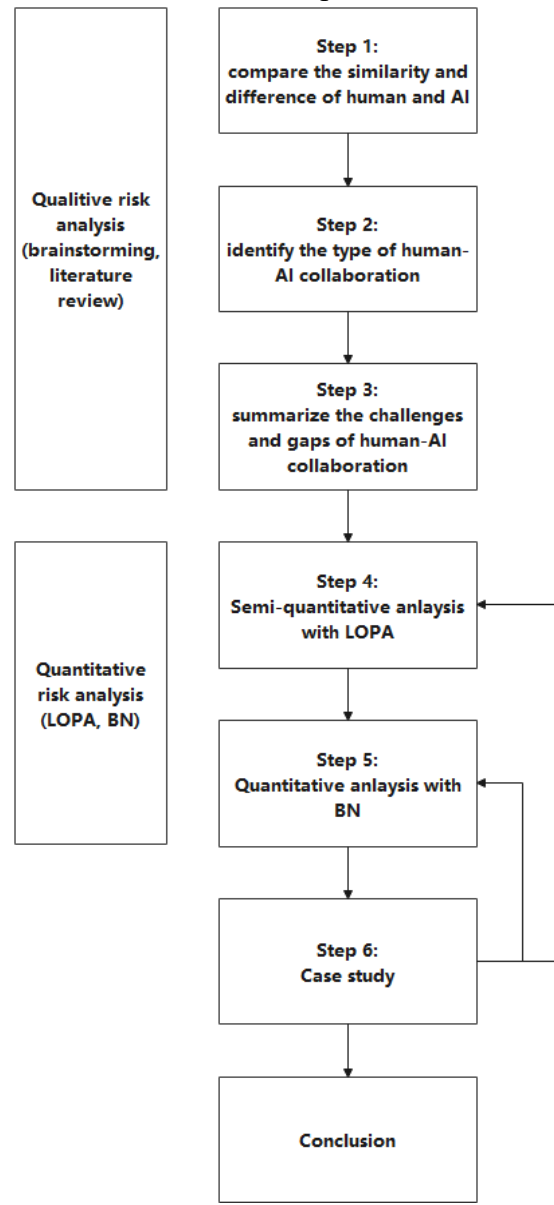

Figure 3: Research flow chart.

Based on the proposed methods and case study, the following questions were discussed: 1) Which is more contributive or effective to process safety, AI or human? 2) What are the traps to avoid accidents when considering increasing the digitalization level or human reliability? 3) How can the uncertainty in failure rate and human error probability be evaluated? The conclusion should be drawn on the paradox of human-AI collaboration and conflict and how to balance the two paths to achieve process safety. The methodology allows to minimize the expert judgment in the definition of human error probability, assess the uncertainty and variability of human errors under different scenarios, investigate human performance under threats and critical state and the interaction between human and AI control systems, model and quantify human reliability with uncertainty.

## 2.2. Key techniques

### 2.2.1. LOPA

LOPA is a semi-quantitative risk assessment method proposed by CCPS (CCPS, 1993), and widely used in process industries, especially in chemical engineering with AI control systems. The preventive layers and mitigative layers consist of the independent protection layers (IPLs) (Figure 1). The frequency of the consequence is computed with the combination of the initial event failure (IEF) and the failure on demand (PFD) of IPL (Table 1).

$$P_{consequence} = IEF \prod_i^n PDF_i \qquad (1)$$

Where $i$ is the number of IPLs.

Table 1: Typical PFDs of IPLs.

| IPL | PFD |
|---|---|
| Control loop failure | 1.00E-02 |
| Human error (trained, no stress) | 1.00E-02 |
| Operator response to alarms | 1.00E-01 |

### 2.2.2. BN

A Bayesian network (BN) is a probabilistic graphical model based on Bayes theorem with a directed acyclic graph (DAG). It relies on the conditional dependence structure of random variables. The probabilities of variables are based on their own frequency observation or accessed in industry handbooks. The conditional probability table (CPT) is usually based on expert knowledge.

$$P(A|B) = \frac{P(B|A)P(A)}{P(B)} \qquad (2)$$

An extension of BN is the credal network (CN), which changes the probability number to a range, as data uncertainty often interferes with the probability of variables.

### 2.2.3. IDHEAS

IDHEAS-Data is the most advanced HRA method, providing specific human error probability simulated by SACADA. IDHEAS-Data (Table 2) contains 4 contexts with 20 performance-influencing factors (PIFs), 133 attributes, and 5 actions with 28 sub-actions. Combining the attributes and sub-actions, it consists of 555 combinations with human error probability, and some combinations have more than one scenario.

Table 2: IDHEAS-Data.

| Context | PIF |
|---|---|
| Environment and situation | Workplace Location Accessibility and Habitability |
| | Workplace Visibility |
| | Noise in Workplace and Communication Pathways |
| | Cold/Heat/Humidity |
| | Resistance to Physical Movement |
| System | System and Instrumentation and Control (I&C) Transparency to Personnel |
| | Human-System Interface (HSI) |
| | Equipment and Tools |
| Personnel | Staffing |
| | Procedures, Guidelines, and Instructions |
| | Training |
| | Team and Organization Factors |
| | Work Processes |
| Task | Information Availability and Reliability |
| | Scenario Familiarity |
| | Multi-Tasking, Interruptions, and Distractions |
| | Task Complexity |
| | Mental Fatigue |
| | Time Pressure and Stress |
| | Physical Demands |

The process of IDHEAS (Xing & Chang, 2018) is:
1) Analyze scenario context.
2) Identify and define human failure events.
3) Analyze tasks and identify critical tasks.
4) Analyze time uncertainty.
5) Analyze cognition failure (identify cognitive failure modes, assess PIFs, and estimate human error probabilities).
6) Adjust HEP based on dependency.
7) Analyze uncertainties.

## 3. Results and discussion

### 3.1. Qualitative risk analysis

#### 3.1.1. Comparison of human and AI in process safety

Based on the AI definition, with brainstorming and literature review, three types or levels of AI in process safety context were identified (Figure 4):

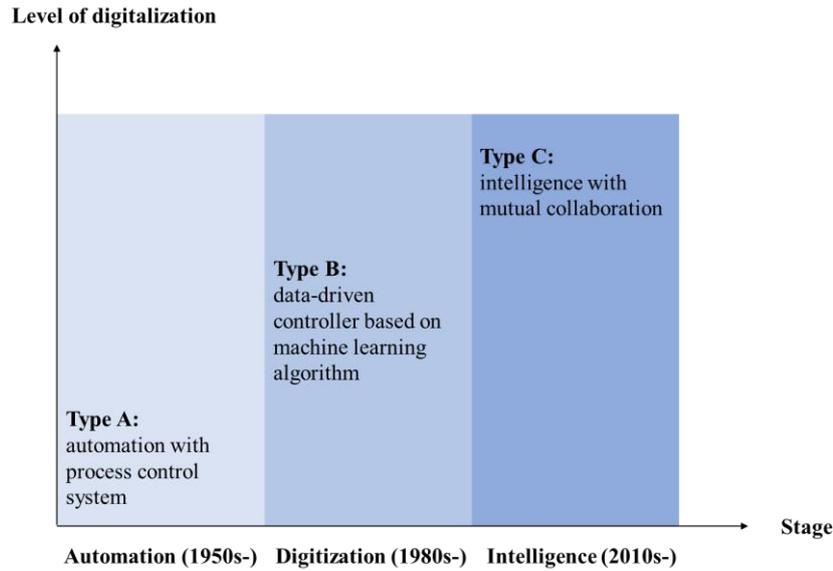

Figure 4: Types of AI in process safety.

1) Type A: automation with the process control system.
   This is the imitation of human control, so-called the AI control system, which at least consists of BPCS, AS, SIS, and ESD (Figure 1). The AI control system is a mature automation application in LOPA. Similarly, according to the automation level definition in IEC 62264 (International Electronichal Commission, 2015), AI control system falls the automation level 1 sensing and manipulating the physical process (sensors, actuators, etc.), and automation level 2 monitoring and controlling of the physical process (programmable logic controller (PLC), distributed control system (DCS), etc.). This type is the dominant application in process industries and the focus of this study.
2) Type B: data-driven controller based on machine learning (ML) algorithm.
   Data-driven controllers are based on ML algorithms, such as model predictive controllers (MPC) and advanced ML models. While a data-driven controller is more likely a black box, learning and predicting from the historical data to export a result, the industrial application is still in a trial due to the high-reliability requirement in process industries. With the development of explainable AI, there will be more practices in the future.
3) Type C: intelligence with mutual collaboration.
   This would be the foreseeable type of AI. Before AI replaces all human labor, humans will work with or alongside AI for a long time. Mutual collaboration means AI can understand human action and response or collaborate correctly for process safety goals.

The significant similarity between AI and humans is that AI imitates human cognition and behavior patterns (Figure 5), following the same procedure to observe, interpret, make decisions, and execute. The difference between AI and humans in process safety is shown in Table 3.

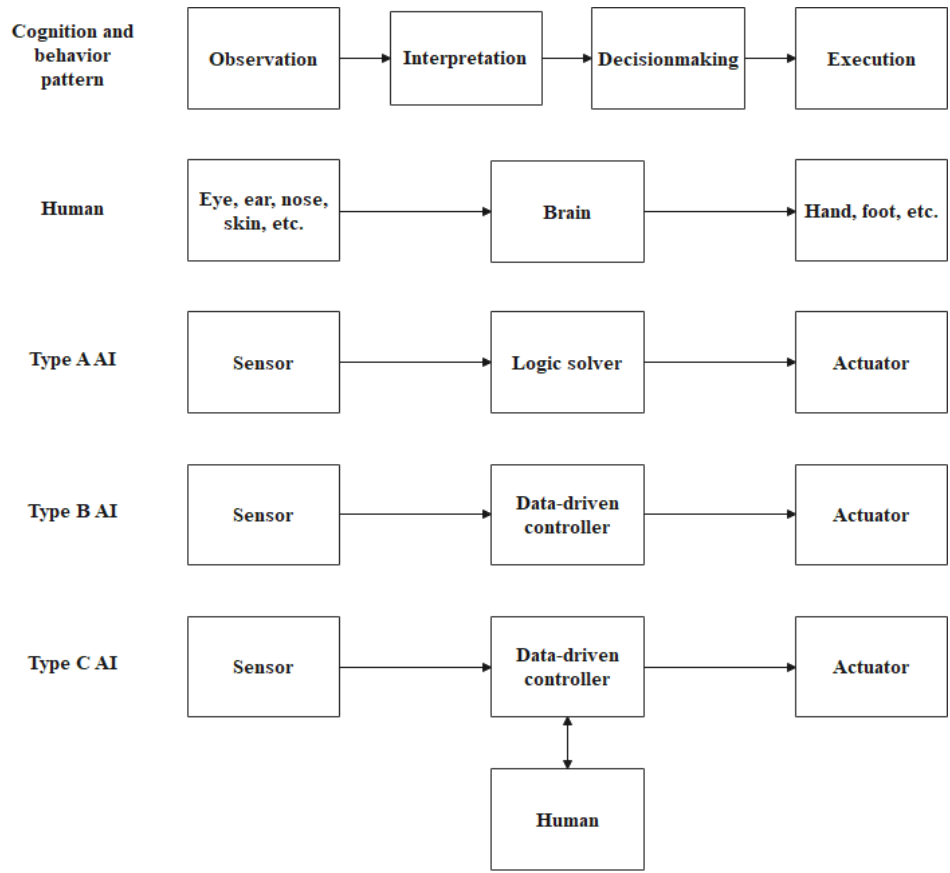

Figure 5: Human and AI working procedure.

Table 3: Difference between humans and AI in the control system.

| Procedure | Human | AI |
|---|---|---|
| Observation | Low precision<br>Qualitative<br>Various objects | High precision<br>Quantitative<br>Fixed objects |
| Interpretation and decision making | Comprehensive understanding<br>Based on memory, experience<br>Learn from failure<br>Flexible<br>Value judged<br>Emotional | Programmed logic<br>Based on data<br>Learn from data<br>Fixed<br>Permanent objective<br>Rational |
| Execution | Sluggish<br>Low precision<br>Fatigue | Agile<br>High precision<br>Low depreciation |

## 3.1.2. Human-AI collaboration AI in process safety

The relation of human-AI collaboration in process safety could refer to similar areas, such as autonomous vehicles, and collaborative robots.

According to the white paper titled "Safety First for Automated Driving" (Aptiv et al., 2019), published by eleven industry leaders across the automotive and automated driving technology spectrum, the levels of driving automation are defined as:

1) level 0-no automation
2) level 1-driver assistance
3) level 2-partial automation
4) level 3-conditional automation
5) level 4-high automation
6) level 5-full automation.

In the research of collaborative robots, the classification of human-robot collaboration was (Ferreira et al., 2021):
1) type 1-coexistence
2) type 2-synchronized collaboration
3) type 3-cooperation
4) type 4 responsive collaboration.

For human-AI collaboration in the control system, the relation will be from assisted cooperation to responsive collaboration, even mutual collaboration (Type C of AI) (Table 4).

Table 4: human-AI collaboration classification.

| Level | Collaboration level | Description | Stage |
| --- | --- | --- | --- |
| Level 3 | Mutual collaboration | AI control system can perform all tasks and interacts with human correctly. | Intelligence stage |
| Level 2 | Responsive collaboration | AI control systems can perform the task by learning from environmental changes. Human monitors only intervene for safety goal. | Digitization stage |
| Level 1 | Assisted cooperation | AI control system works with a structured program, and humans are assistive to monitor and intervene at any time. | Automation stage |
| Level 0 | Manual control | The human controls manually. | - |

The current stage is in level 1, developing to level 2, and the work process of human-AI collaboration is shown in Figure 6. At first, the AI control system works for the system goal, and human assists as a standby component. Once AI fails, human intervenes to bring it back to balance or ensure the safety goal. Human usually follows AI in this sequential relation.

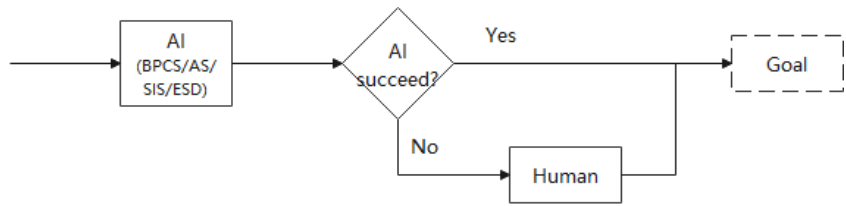

Figure 6: Sequential relation between humans and AI.

### 3.1.3. Challenge and gap analysis of human-AI collaboration

Based on the study of the classification, difference, and work process of human-AI collaboration, the challenges and gaps of human-AI collaboration were summarized (Table 5):

1) Data and information flow

    The sensor generates structured data, such as temperature, pressure, level, etc., which contain limited information just for some objective. However, humans could receive or obtain different sources of data and information. The data output of AI and human input are both limited. The sensor data may be uncertain, imperfect, sometimes inherently biased, or even dirty with inappropriate feedback and miscellaneous information. It makes the AI model biased at the start.

2) Algorithm and model

    A controller is usually an integrated product with a fixed program and solid function. As the users, the operators are required little ability or a chance to modify it. For the fuzzy algorithm and model, spurious interruption or execution may result in unreasonable phenomena. In other words, once deviations happen, the operators cannot fix them on-site. However, the tolerance expectation for AI failure is lower than human error. Even worse, AI is more vulnerable to noise or specific inputs when deployed, which are so-called adversarial attacks.

3) Operation and maintenance

    On the opposite of low dependability to operators, AI control system relies on IT operations heavily, while IT support may not be available timely. On the other hand, as is known, cybersecurity issues are increasing with the development all the time.

4) Work division and priority

    Apparently, if AI dominates the human-AI control system, it would be a disaster as AI has no value judgment on human life and loss prevention. Otherwise, from the perspective of inherent safety, AI tends to be foolproof and rectification on human error; once AI sets human corrective action to be human error, AI will only execute the fixed program ignoring the circumstances. This is the case between the pilot and Maneuvering Characteristics Augmentation System (MCAS) in Boeing 737 Max accidents (The House Committee on Transportation and Infrastructure, 2020).

5) Potential threats by human

    Correct actions are limited, while unpredicted human actions are various. In contrast, the current AI control system cannot adapt or respond to such deviated human interactions. Moreover, in the digital age, humans could access the system anywhere at any time if authorized.

6) Environment

    As talked about above, AI has low adaptability to dynamic environments, especially in emergencies. Sometimes neither does human. The challenges from circumstance changes are the critical factor to human-AI collaboration.

7) Unknown factors and consequences

    In system initiation and operation, unforeseeable new problems may show up due to unpredictable reasons. Humans and AI are both challenged to face such

unknowns.

Table 5: Challenge and gap of human-AI collaboration.

| Item | Challenge |
|---|---|
| Data and information flow | Structured data with limited information |
| | Uncertain, imperfect, biased data |
| | Inappropriate feedback data |
| | Limited AI output and human input |
| Algorithm and model | Fixed program with low changeability by operators |
| | Spurious interruption or execution |
| | Lower tolerance expectation on AI failure than human |
| | adversarial attacks |
| Operation and maintenance | High dependency on IT operations |
| | Increasing cybersecurity issue |
| Work division and priority | Dominance in the work division |
| | Priority paradox between foolproof and rectification |
| Potential threats by human | Unpredictable human interactions |
| | Connected humans with authority anywhere |
| Environment | Low adaptability of AI for a dynamic environment |
| Unknown | Unforeseeable new problems |

## 3.2. Quantitative risk analysis

### 3.2.1. LOPA analysis

LOPA is applied to evaluate the human-AI collaboration. The IPLs in the scenario of human-AI collaboration are BPCS, alarm with operator intervention, SIS, and manual shutdown action. A universal pattern of human-AI collaboration in LOPA is shown in Figure 7.

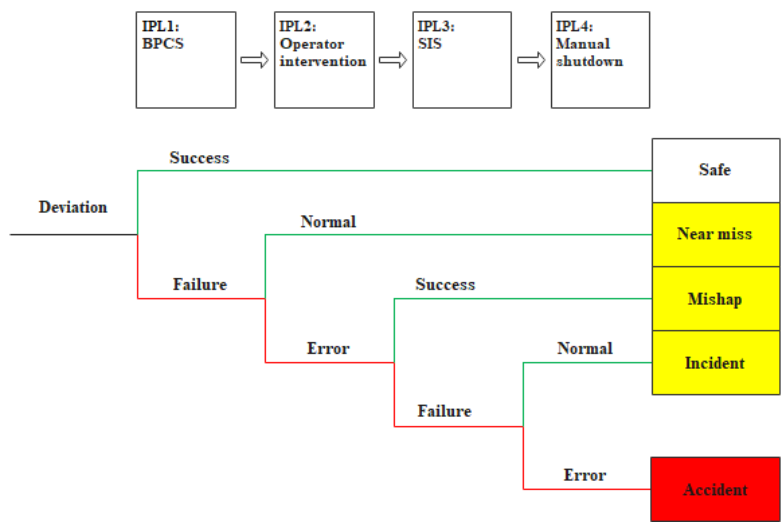

Figure 7: AI-human collaboration by LOPA.

The initial event is any deviation in the control system, which exists all the time, considering the deviation frequency as 100%. The undesired but tolerable consequences are near miss (BPCS failure), mishap (SIS initiation), and incident (manual shutdown success). The accident is to fail to manually shutdown, and the probability is:

$$P_{accident} = \prod_i^4 PDF_i \tag{3}$$

Where $i$ is the number of IPLs.

It seems that to prevent an undesired event, AI plays the same role as humans; it is the same to increase AI layers or human actions. However, increasing human actions may have counter effects because human action is attached to the failure of each AI failure, as shown above in Figure 6.

Another way is to increase the reliability of each AI layer and each human action. In some industry practices, this principle guides the designer to chase the digitalization level, ignoring the human action attached to AI failure. As there is much more space to enhance human reliability than AI, the focus needs to fall on humans, which is why human-centered AI (Shneiderman, 2021).

### 3.2.2. BN analysis

BN is applied to provide specific dependency and probability analysis to improve QRA of human-AI collaboration.

In the human-AI collaboration, AI failures were identified as systematic, sensor, logic solver, and actuator failures. Specifically, systematic failures include system integration, network communication, power supply outage, and I/O (input/output) connection failure. The logic solver failures consist of model uncertainty, the control logic unit failure, storage insufficient, and random disturbance. The failure data of components could be found in OREDA with lower and upper limits, and others could be estimated with expert knowledge and industry practice.

This study only used the PIFs, and the HEPs of attributes and combinations, not strictly following the whole calculation process of IDHEAS-Data. According to IDHEAS-Data, the 20 PIFs should be all considered, while the following 15 are more significant in human-AI collaboration: noise, heat/cold, system transparency, human-system interface, equipment and tools, staffing, procedures, training, team and organization, work processes, information availability, scenario familiarity, task complexity, mental fatigue, time pressure, and stress.

Based on the dependency analysis and cause analysis, the BN of human-AI collaboration was constructed (Figure 8).

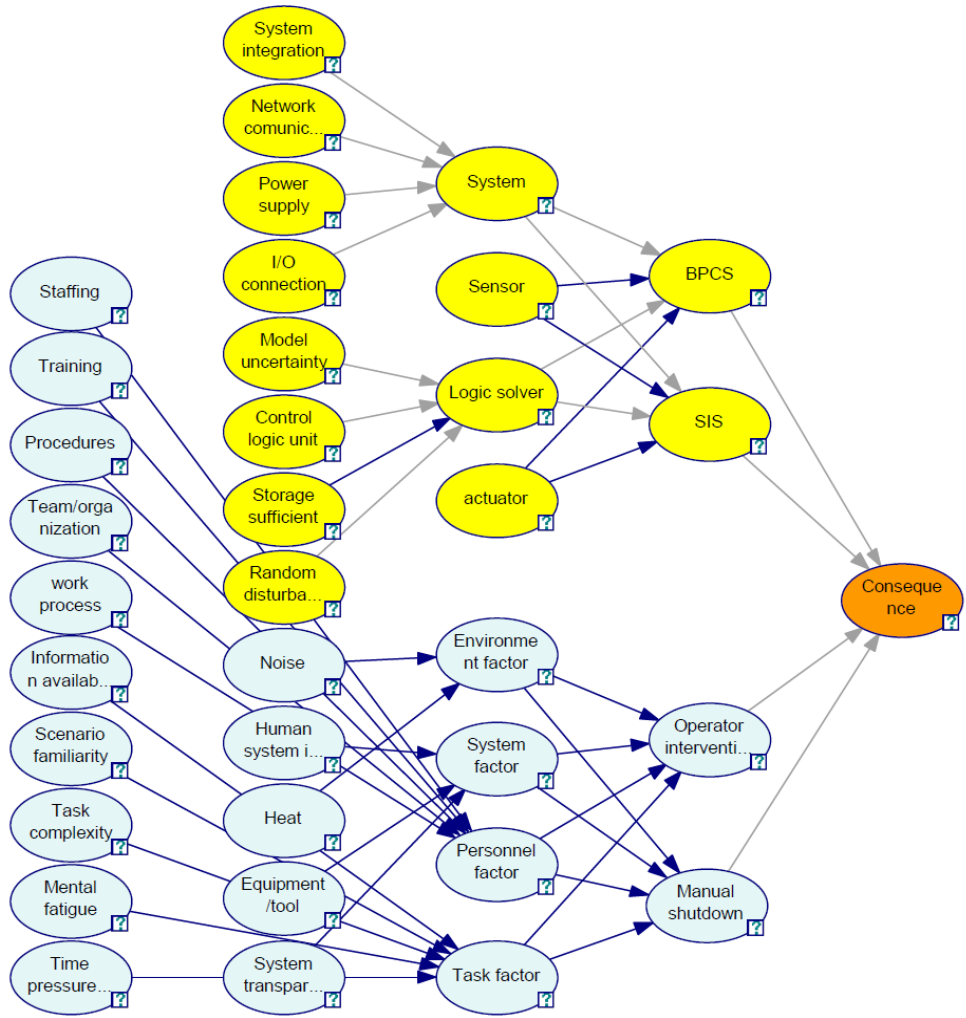

Figure 8: BN analysis of human-AI collaboration.

The HEP should be a range as human performance varies in dynamic circumstances and personal conditions. So does the physical component. The probability range will depend on data uncertainty. As all probabilities in this study are based on experience accumulated in practice, and they all provide the lower probability limit (LPL) and upper probability limit (UPL). The scenarios can be set as the best and the worst cases. The credal network is an extension of the Bayesian network, using a range instead of a fixed number.

### 3.3. Case study

#### 3.3.1. Case description

This study selected a published case (Thepmanee, 2018) of a two-phase gas-liquid separator process (Figure 9) on integrated BPCS and SIS System reliability modeling.

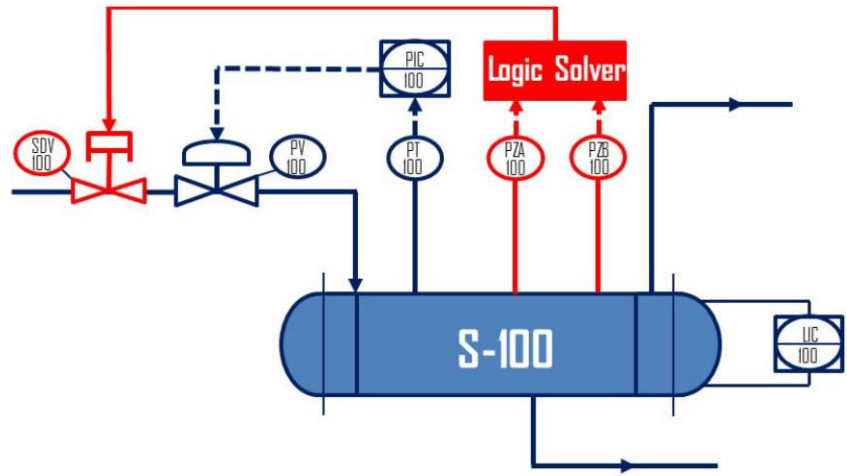

Figure 9: Two-phase gas-liquid separator process (Thepmanee, 2018).

The process introduction from this article was written as follows:
1) The pressure transmitter (PT-100) measures the pressure of the separator and sends a signal to the pressure controller. The pressure control valve (PV-100) is a final element of the control system that regulates the flow of the inlet feed stream. The pressure controller (PIC-100) processes the signal received from the PT-100 and then actuates a pressure control valve PV-100. The inlet feed stream is controlled by PV-100 valve to control the pressure of the separator. PIC-100 determined this valve.
2) It is assumed that the separator's pressure is designed to operate at a lower than the pressure of the inlet feed. When the PV-100 valve lacks control, the inlet stream supplies surplus feed. This leads to the separator's overpressure condition and is possible to cause an unpleasant explosion. To avoid this accident, ESD should be therefore prepared.

This article applied fault tree analysis (Figure 10) with the given failure data (Table 6) and did not consider any human intervention. The article simulated the dynamic change of system failure probability in one year without giving specific numbers. Basically, in this method, the failure of the process in one year should be 8.15E-04.

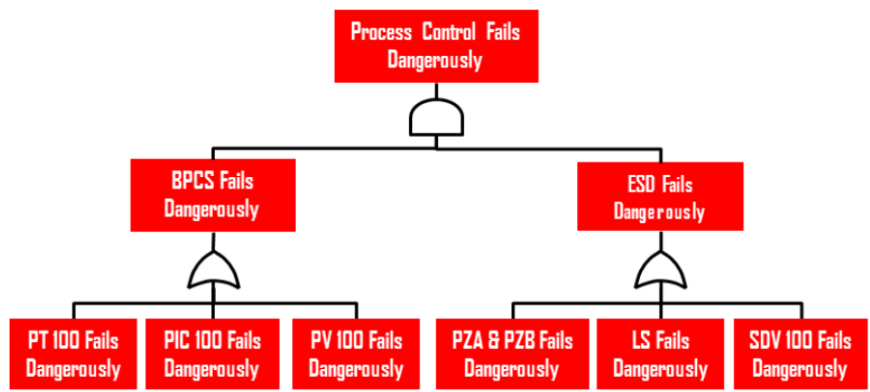

Figure 10: Fault tree analysis (Thepmanee, 2018).

Table 6: Failure data of each component.

| Component | Failure rate in 1.00E-06 hours | Failure probability in one year |
|---|---|---|
| PT-100 | 500 | 1.25E-02 |
| PIC-100 | 800 | 9.05E-04 |
| PV-100 | 1500 | 1.97E-06 |
| PZA&PZB | 346 | 4.83E-02 |
| Logic solver | 500 | 1.25E-02 |
| SDV-100 | 1340 | 7.98E-06 |

### 3.3.2. LOPA analysis for case study

As the case article did not consider human actions, according to LOPA described in subsection 3.2.1, the probabilities of IPLs were assumed based on Table 1:
1) IPL1 BPCS: 1.00E-02.
2) IPL2 operator intervention: 1.00E-01.
3) IPL3 SIS: 1.00E-02.
4) IPL4 manual shutdown: 1.00E-02.

According to equation (1), the probability of an accident is 1.00E-07.

### 3.3.3. BN analysis for case study

According to the case description, the probability of AI failure (Table 7) was obtained from OREDA and expert knowledge, and HEPs (Table 8) could be identified in IDHEAS-Data. Based on expert knowledge, the CPTs could be determined, and one example of the CPTs was shown in Table 9. Figure 11 is the BN analysis result based on the best scenario with all lower probability limits.

Table 7: AI failure probabilities.

| Failure type | LPL | UPL |
|---|---|---|
| Sensor failure | 1.05E-01 | 2.15E-01 |
| Actuator | 1.05E-01 | 2.15E-01 |
| System integration failure | 1.00E-02 | 2.00E-02 |
| Network communication failure | 2.69E-02 | 5.63E-02 |
| Power supply failure | 7.54E-02 | 7.54E-02 |
| I/O connection failure | 2.69E-02 | 5.63E-02 |
| Model uncertainty | 2.69E-02 | 5.63E-02 |
| Control logic unit (CLU) failure | 4.73E-02 | 4.73E-02 |
| Storage insufficient | 2.69E-02 | 5.63E-02 |
| Random disturbance beyond control | 3.40E-06 | 3.40E-06 |

Table 8: Human error probabilities.

| Context | PIF | LPL | UPL |
|---|---|---|---|
| Environment | Noise | 2.10E-02 | 2.80E-02 |
|  | Heat | 1.00E-01 | 2.00E-01 |
| System | System Transparency | 3.00E-02 | 1.50E-01 |
|  | Human-System Interface | 4.00E-03 | 1.40E-02 |
|  | Equipment and Tools | 5.20E-02 | 7.20E-02 |
| Personnel | Staffing | 4.80E-02 | 9.80E-02 |
|  | Procedures | 3.30E-02 | 6.90E-02 |
|  | Training | 3.60E-02 | 4.50E-02 |
|  | Team and Organization | 1.00E-01 | 1.60E-01 |
|  | Work Processes | 7.00E-02 | 1.10E-01 |
| Task | Information Availability | 3.00E-02 | 2.80E-01 |
|  | Scenario Familiarity | 1.40E-02 | 1.70E-01 |
|  | Task Complexity | 2.10E-03 | 1.56E-02 |
|  | Mental Fatigue | 2.00E-01 | 3.00E-01 |
|  | Time Pressure and Stress | 5.62E-02 | 3.50E-01 |

Table 9: One example of the CPTs.

| Noise | TRUE | | FALSE | |
|---|---|---|---|---|
| Heat | TRUE | FALSE | TRUE | FALSE |
| TRUE | 1.00E-00 | 2.10E-02 | 1.00E-01 | 0.00E-00 |
| FALSE | 0.00E-00 | 9.79E-01 | 9.00E-01 | 1.00E-00 |

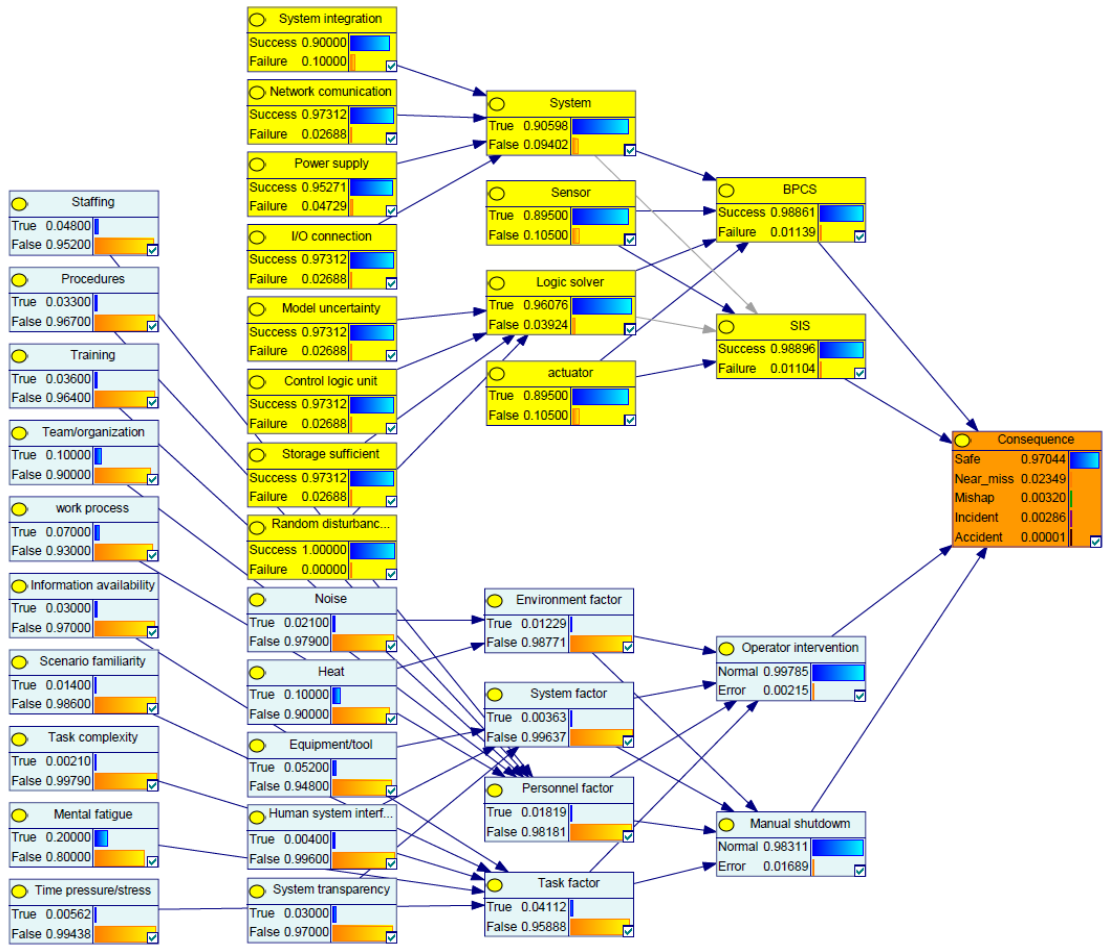

Figure 11: BN analysis results with the best scenario.

To change the data with upper probabilities in Table 7 and Table 8, then repeat the BN analysis, the worst scenario was achieved (Table 10).

Table 10: The best and worst scenarios.

| Consequence | LPL | UPL |
|---|---|---|
| Safe | 9.70E-01 | 9.63e-01 |
| Near miss | 2.35E-02 | 3.17e-02 |
| Mishap | 3.20E-03 | 3.94e-03 |
| Incident | 2.86E-03 | 1.19e-03 |
| Accident | 7.62E-06 | 1.10e-04 |

Comparing the two scenarios (Figure 12), the similarity is that near miss occupies the largest share in the unsafe conditions (near miss, mishap, incident, accident), and it implies that to prevent the undesired consequence as early as possible. In other words, increasing the reliability of BPCS and the practical action by the operator is vital to the system.

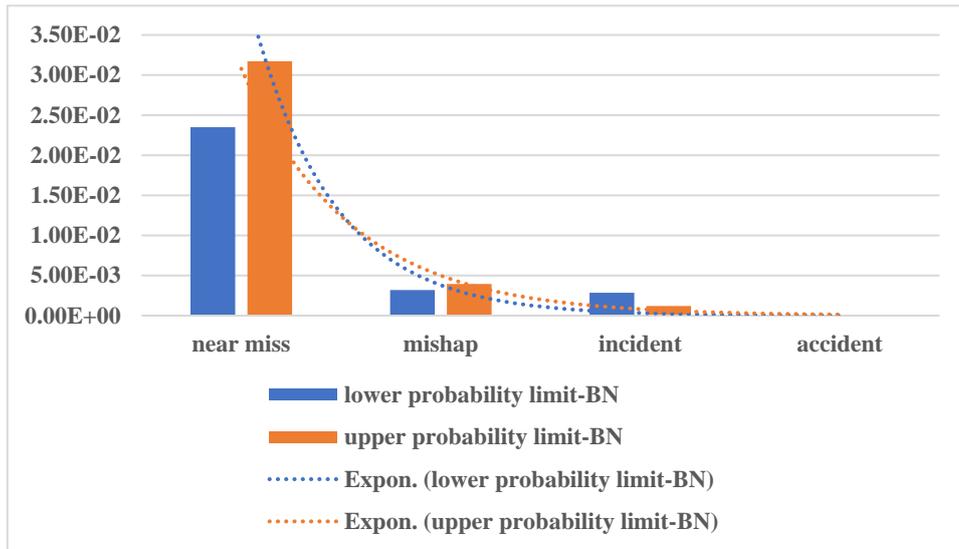

Figure 12: Comparison of the unsafe conditions.

Comparing the accident probability of the original case, LOPA result, and BN result (Figure 13), human participation is a significant way to compensate AI failure, even with rough calculation method (LOPA) with reasonable estimated probabilities. Furthermore, BN could specify human factors and AI failure causes, provide more precise calculation, and the results keep the reasonable variance.

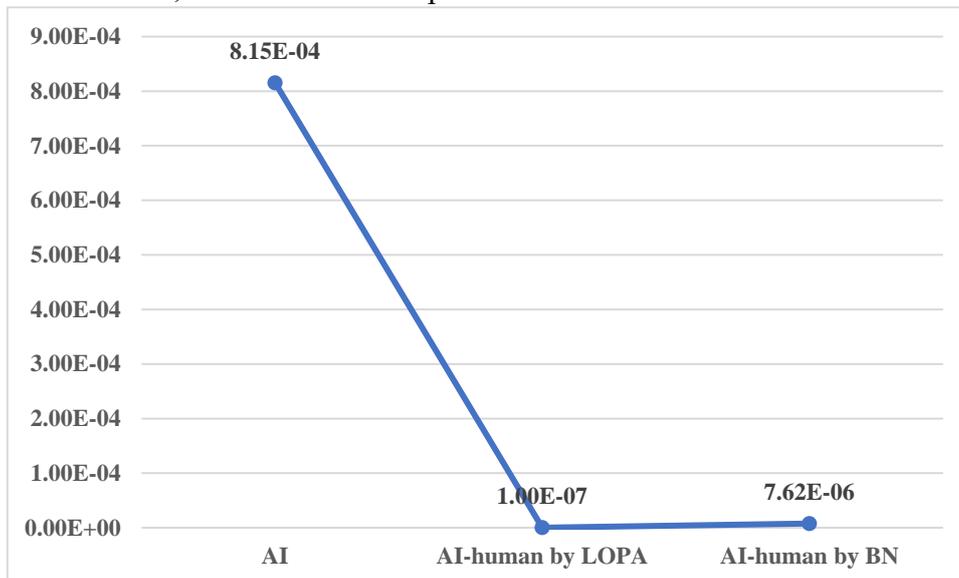

Figure 13: Accident probability comparison.

## 4. Conclusion

In conclusion, this study identified the types of AI in process safety, summarized the difference between humans and AI, interpreted how humans and AI collaborate, and classified the human-AI collaboration. It contributed to the inadequate research on digitalization's impact on process safety.

Furthermore, the challenges and gaps were analyzed with qualitative thinking. And the risk in human-AI collaboration was assessed quantitatively with LOPA and BN. Significantly, the BN presented the anatomy of the fundamental causes of AI failure, and the primary factor of human error, bringing the two parts together, not separated analysis as before. Some new ignored causes were proposed and analyzed, such as random variable-caused failure, system integration failure, and model uncertainty.

The case study illustrated the proposed methods, addressing the result of quantitative benchmarking in different scenarios. It stressed the impact of the data uncertainty in AI failure rate and HEP on accident probability. Also, the discussion on the effective way to achieve process safety was made, and the key finding was that human is still the most critical but unpredictable, unreliable, and vulnerable part in human-AI system; it needs attention to the trap of improving the automation capability and digitalization level continuously, ignoring the human error.

The next phase of digitalization may push the human-AI collaboration from assisted collaboration to responsive collaboration, even mutually responsive collaboration. However, it should be a long way to go, and before that, the most influencing method is to draw attention to human error.

The limitation of the work is that there is no further exploration of ML-based controllers, and mutual intelligent control, as this is still in the trial phase. And some probabilities applied in the research were still from human judgment, though they are based on expert knowledge. More precise possibilities should be considered; before that, an imprecise probability range can only be used for comparative study. The industry application will undoubtedly accumulate more correct and suitable data on AI failure and human error.

**Reference**

Aaltonen, I., Salmi, T., & Marstio, I. (2018). Refining levels of collaboration to support the design and evaluation of human-robot interaction in the manufacturing industry. *Procedia CIRP*, *72*, 93–98. https://doi.org/10.1016/j.procir.2018.03.214

Alizadeh, S., & Sriramula, S. (2018). Unavailability assessment of redundant safety instrumented systems subject to process demand. *Reliability Engineering and System Safety*, *171*(December 2016), 18–33. https://doi.org/10.1016/j.ress.2017.11.011

Aptiv, Audi, Baidu, BMW, Continental, FCA, Here, Infineon, Intel, & Volkswagen. (2019). Safety first for automated driving 2019. In *White paper of different car manufacutres and suppliers*. https://www.press.bmwgroup.com/global/article/attachment/T0298103EN/434404

Baybutt, P. (2002). Layers of protection analysis for human factors (LOPA-HF). *Process Safety Progress*, *21*(2), 119–129. https://doi.org/10.1002/prs.680210208

Benson, C., Argyropoulos, C. D., Dimopoulos, C., Mikellidou, C. V., & Boustras, G. (2021). Safety and risk analysis in digitalized process operations warning of possible deviating conditions in the process environment. *Process Safety and Environmental Protection*, *149*, 750–757.

https://doi.org/10.1016/j.psep.2021.02.039

Briken, K. (2020). Welcome in the machine: Human-machine relations and knowledge capture. *Capital and Class*, *44*(2), 159–171. https://doi.org/10.1177/0309816819899418

Cadwallader, L. C. (1996). Reliability Estimates for Selected Sensors in Fusion Applications. *Contract*, *September*.

CCPS. (1993). *Guidelines for safe automation of chemical processes*. Center for chemical process safety of the American Institute of chemical ….

Chang, K., Kim, S., Chang, D., Ahn, J., & Zio, E. (2015). Uncertainty analysis for target SIL determination in the offshore industry. *Journal of Loss Prevention in the Process Industries*, *34*, 151–162. https://doi.org/10.1016/j.jlp.2015.01.030

Chen, X., Despeisse, M., & Johansson, B. (2020). Environmental sustainability of digitalization in manufacturing: A review. *Sustainability (Switzerland)*, *12*(24), 1–33. https://doi.org/10.3390/su122410298

Cohen, Y., & Singer, G. (2021). A smart process controller framework for Industry 4.0 settings. *Journal of Intelligent Manufacturing*, *32*(7), 1975–1995. https://doi.org/10.1007/s10845-021-01748-5

Dowell, A. M. (1998). Layer of protection analysis for determining safety integrity level. *ISA Transactions*, *37*(3), 155–165. https://doi.org/10.1016/s0019-0578(98)00018-4

Edwards, E., & Lees, F. P. (1971). The development of the role of the human operator in process control. *IFAC Proceedings Volumes*, *4*(3), 138–144. https://doi.org/10.1016/s1474-6670(17)68589-6

European Agency for Safety & Health at Work [EU-OSHA]. (2018). *Foresight on new and emerging occupational safety and health risks associated with information and communication technologies and work location by 2025*. https://osha.europa.eu/pt/publications/foresight-new-and-emerging-occupational-safety-and-health-risks-associated

Ferreira, C., Figueira, G., & Amorim, P. (2021). Scheduling human-robot teams in collaborative working cells. *International Journal of Production Economics*, *235*(May 2020), 108094. https://doi.org/10.1016/j.ijpe.2021.108094

Forester, J., Kolaczkowski, A., Lois, E., & Kelly, D. (2006). NUREG-1842: Evaluation of Human Reliability Analysis Methods Against Good Practices. In *U.S. Nuclear Regulatory Commission Office of Nuclear Regulatory Research Washington, DC 20555-0001*.

Freschi, C., Ferrari, V., Melfi, F., Ferrari, M., Mosca, F., & Cuschieri1, A. (2013). Technical review of the da Vinci surgical telemanipulator. *The International Journal of Medical Robotics and Computer Assisted Surgery*, *9*, 396–406. https://doi.org/10.1002/rcs.1468

Ghobakhloo, M. (2020). Industry 4.0, digitization, and opportunities for sustainability. *Journal of Cleaner Production*, *252*, 119869. https://doi.org/10.1016/j.jclepro.2019.119869

Hollnagel, E. (1998). *Cognitive reliability and error analysis method (CREAM)*. Elsevier.

Hollnagel, E. (2018). *Safety-I and safety-II: the past and future of safety management*. CRC press.

Huang, J., Li, Z., & Liu, H.-C. (2017). New approach for failure mode and effect analysis using linguistic distribution assessments and TODIM method. *Reliability Engineering & System Safety*, *167*, 302–309. https://doi.org/https://doi.org/10.1016/j.ress.2017.06.014

International Electronichal Commission. (2015). Functional safety: What is Functional safety ? *International Electrotechical Commission*. https://www.iec.ch/resource-centre/functional-safety-essential-overall-safety

Ivanov, D., Dolgui, A., & Sokolov, B. (2019). The impact of digital technology and Industry 4.0 on the ripple effect and supply chain risk analytics. *International Journal of Production Research*, *57*(3), 829–846. https://doi.org/10.1080/00207543.2018.1488086

Khan, F., Amyotte, P., & Adedigba, S. (2021). Process safety concerns in process system digitalization. *Education for Chemical Engineers*, *34*, 33–46. https://doi.org/10.1016/j.ece.2020.11.002

Kirwan, B. (1998). Human error identification techniques for risk assessment of high risk systems - Part 1: Review and evaluation of techniques. In *Applied Ergonomics* (Vol. 29, Issue 3, pp. 157–177). https://doi.org/10.1016/S0003-6870(98)00010-6

Lee, J., Cameron, I., & Hassall, M. (2019). Improving process safety: What roles for digitalization and industry 4.0? *Process Safety and Environmental Protection*, *132*, 325–339. https://doi.org/10.1016/j.psep.2019.10.021

Liu, Y., & Rausand, M. (2013). Reliability effects of test strategies on safety-instrumented systems in different demand modes. *Reliability Engineering and System Safety*, *119*, 235–243. https://doi.org/10.1016/j.ress.2013.06.035

Magrini, E., Ferraguti, F., Ronga, A. J., Pini, F., De Luca, A., & Leali, F. (2020). Human-robot coexistence and interaction in open industrial cells. *Robotics and Computer-Integrated Manufacturing*, *61*(June 2018), 101846. https://doi.org/10.1016/j.rcim.2019.101846

McCluskey, P. (2002). Design for reliability of micro-electro-mechanical systems (MEMS). In *Proceedings - Electronic Components and Technology Conference* (pp. 760–762). https://doi.org/10.1109/ectc.2002.1008183

Myers, P. M. (2013). Layer of protection analysis - Quantifying human performance in initiating events and independent protection layers. *Journal of Loss Prevention in the Process Industries*, *26*(3), 534–546. https://doi.org/10.1016/j.jlp.2012.07.003

Nasimi, E., & Gabbar, H. A. (2016). Application of Safety Instrumented System (SIS) approach in older nuclear power plants. *Nuclear Engineering and Design*, *301*, 1–14. https://doi.org/10.1016/j.nucengdes.2016.02.026

OREDA. (2021). *History of OREDA*. https://www.oreda.com/history

Park, J., Jeon, H., Kim, J., Kim, N., Park, S. K., Lee, S., & Lee, Y. S. (2019). Remaining and emerging issues pertaining to the human reliability analysis of domestic nuclear power plants. *Nuclear Engineering and Technology*, *51*(5),

1297–1306. https://doi.org/10.1016/j.net.2019.02.015

Riccardo, P., Marilia, R., Paltrinieri, N., Massaiu, S., Costantino, F., Di Gravio, G., & Boring, R. L. (2020). Human reliability analysis: Exploring the intellectual structure of a research field. In *Reliability Engineering and System Safety* (Vol. 203). https://doi.org/10.1016/j.ress.2020.107102

Shneiderman, B. (2021). Human-centered AI. *Issues in Science and Technology*, *37*(2), 56–61.

SINTEF. (2021). *Reliability Data for Safety Instrumented Systems: PDS Data Handbook*. SINTEF, Technology and Society, Safety and Reliability. https://books.google.ca/books?id=Emc5OgAACAAJ

Tadigadapa, S., & Najafi, N. (2001). Reliability of micro-electro-mechanical systems (MEMS). *Reliability, Testing, and Characterization of MEMS/MOEMS*, *4558*, 197–205.

The House Committee on Transportation and Infrastructure. (2020). *Final commette Report: The Design, Development & Certification of the Boeing 737 Max* (Issue September).

Thepmanee, T. (2018). Reliability modeling for integrated BPCS and ESD system case study: A two-phase gas-liquid separator process. *2018 3rd International Conference on Control and Robotics Engineering, ICCRE 2018*, 165–168. https://doi.org/10.1109/ICCRE.2018.8376456

USNRC. (1975). WASH-1400 (NUREG-75/014): Reactor safety study: An assessment of accident risks in U.S. commercial nuclear power plants. In *WASH-1400 (NUREG-75/014): Reactor safety study: An assessment of accident risks in U.S. commercial nuclear power plants* (Issue October). http://www.nrc.gov/reading-rm/doc-collections/nuregs/staff/sr75-014/

Wang, P. (2008). What Do You Mean by " AI "? The Problem of AI Typical Ways to Define AI. *Agi*, *171*, 363–373.

Wu, H. N., Zhang, X. M., & Li, R. G. (2021). Synthesis with guaranteed cost and less human intervention for human-in-the-loop control systems. *IEEE Transactions on Cybernetics*, 1–11. https://doi.org/10.1109/TCYB.2020.3041033

Xing, J., & Chang, Y. J. (2018). Use of IDHEAS general methodology to incorporate human performance data for estimation of human error probabilities. *PSAM 2018 - Probabilistic Safety Assessment and Management*, September.